\pdfoutput=1
\documentclass[preprint]{elsarticle}

\usepackage{graphicx}
\usepackage{amsmath}
\usepackage{amssymb}
\usepackage{color}
\biboptions{numbers,sort&compress}
\title{Lepton Flavour Violation in a radiative neutrino mass model with the asymmetric Yukawa structure}
\author[1]{Yoko Irie}
\author[2,3]{Osamu Seto}
\author[4]{Tetsuo Shindou}
\address[1]{Department of Applied Physics, \\
Kogakuin University, Hachioji, Tokyo, 192-0015, Japan}
\address[2]{Institute for the Advancement of Higher Education,\\
 Hokkaido University, Sapporo 060-0817, Japan}
\address[3]{
Department of Physics, \\
Hokkaido University, Sapporo 060-0810, Japan
}
\address[4]{Division of Liberal-arts, \\
Kogakuin University, Hachioji, Tokyo, 192-0015, Japan}

\begin{document}
\begin{abstract}
Though models with the radiative neutrino mass generation are phenomenologically attractive, the complicated relationship between the flavour structure of additional Yukawa matrices and the neutrino mass matrix sometimes is a barrier to explore the models. We introduce a simple prescription to analyze the relation in a class of models with the asymmetric Yukawa structure. We then apply the treatment to the Zee-Babu model as a concrete example of the class and discuss the phenomenological consequences of the model. The combined studies among the neutrino physics, the lepton flavour violation, and the search for the new particles at the collider experiments provide the anatomy of the Zee-Babu model.
\end{abstract}

\maketitle

\section{Introduction}

The neutrino oscillation experiments show that neutrinos have finite masses.
The masses are several orders of magnitude smaller than the electron mass so that the origin of neutrino masses seems to be different from the other Standard Model (SM) fermions whose masses are generated through the Yukawa couplings with the Higgs boson.
It is a big mystery in the SM what is the origin of such tiny neutrino masses. 

There are many attractive ideas to address the origin of the neutrino mass. 
The most popular one is the canonical seesaw mechanism~\cite{Minkowski:1977sc, Yanagida:1979as, 
Yanagida:1980xy,
GellMann:1980vs,
Mohapatra:1979ia} where heavy sterile neutrinos are introduced into the SM, and the dimension-five operator so-called the Weinberg operator~\cite{Weinberg:1979sa} is induced after the integration of the heavy sterile neutrinos.
The coefficient of the Weinberg operator is suppressed by the very large mass scale of the sterile neutrinos, which is usually taken to be larger than $10^6$~GeV. 
There are another class of models in which the tininess of the neutrino mass is responsible for the tiny vacuum expectation value (VEV) of an extra scalar field. 
A popular realisation of this idea is the type-II seesaw model~\cite{Konetschny:1977bn,Cheng:1980qt,Lazarides:1980nt,Schechter:1980gr,Magg:1980ut,Mohapatra:1980yp} where a $SU(2)$ triplet scalar field develops its tiny VEV. 
If the neutrino Yukawa coupling with the extra scalar field is of the order of $0.1$, the VEV of the extra scalar field is expected to be as small as $1$~eV. 

An alternative approach to the origin of neutrino masses is that the tiny neutrino masses are generated through the quantum loop effect. 
Since the first model of this class was proposed by A.~Zee~\cite{Zee:1980ai}, many interesting models have been proposed.
A comprehensive review of this class of models can be found, for example, in Ref.~\cite{Cai:2017jrq}.
Such models are phenomenologically attractive because the new particles are introduced at around the TeV scale or below. 
A model can be explored through a new particle direct production by future collider experiments as well as significant contributions to the flavour physics by new particles. 

Since the neutrino oscillation parameters become precisely determined by the experiments~\cite{Esteban:2020cvm}, 
the Lagrangian parameters such as Yukawa coupling matrices with the extra scalars are restricted to reproduce the correct neutrino mass matrix. 
However, in many models, the relation between the Lagrangian parameters and the parameters in the induced neutrino mass matrix is not simple. 
The neutrino mass matrix is given by a product of several matrices, including relevant Yukawa coupling matrices, and there are additional degrees of freedom that cannot be determined by the input of the neutrino parameters.
This will be a barrier to explore the models by flavour phenomenology.

Nevertheless, as far as models with loop generated neutrino masses are concerned, a comprehensive classification of the models based on the flavour structure has been completed in Refs.~\cite{Kanemura:2015cca, Kanemura:2016ixx}.
The models which induce Majorana neutrino masses are classified into only three groups, and 
the models with Dirac neutrino masses are classified into seven groups. 
This is a great step to a systematic analysis of the phenomenological feature of the models. 

In this article, we focus on a group of the models with the asymmetric Yukawa structure, to which the Zee-Babu model~\cite{Zee:1985id, Babu:1988ki} and the Krauss-Nasri-Trodden model~\cite{Krauss:2002px} belong as popular examples. 
We propose a convenient treatment of relations between the Lagrangian parameters and the neutrino mass matrix.
Once we find the treatment, it is easy to apply it to a concrete model and to study flavour phenomenology. 

We will then demonstrate how our treatment works in the Zee-Babu model as an example. 
In the Zee-Babu model, a singly charged singlet scalar and a doubly charged singlet scalar are contained, and the neutrino masses are induced at the two-loop.
Rich phenomenology in the model has been widely studied in the literature, for instance, see Refs.~\cite{Babu:2002uu,AristizabalSierra:2006gb,Nebot:2007bc,Schmidt:2014zoa,Herrero-Garcia:2014hfa,Alcaide:2017dcx}.
Utilizing the relation between the Lagrangian parameters and the neutrino mass matrix, 
we can systematically analyze the flavour phenomenology in the model, and we will show the constraint on the CP phases in the neutrino mixing matrix. 

This letter is organized as follows. 
In Sec.~2, we show how to extract the Yukawa matrices in the Lagrangian by the input of the neutrino mass matrix.
In Sec.~3, we apply the treatment to the Zee-Babu model, and we analyze the phenomenological feature of the Zee-Babu model. 
We give Summary in Sec.~4.

\section{Flavour structure}\label{Sec_Flavour_Structure}

Though there are many models with radiative Majorana neutrino mass generation proposed,
 those can be classified into only three groups by their flavour structure~\cite{Kanemura:2015cca}.
We here focus on models of the Group-I in Ref~\cite{Kanemura:2015cca}, in which the light neutrino mass matrix $M_{\nu}$ is given by 
\begin{equation}
M_{\nu}\propto Y_{\omega}m_{\ell}X_Sm_{\ell}Y_{\omega}^T\equiv M\;,	
\label{FlavorStructure}
\end{equation}
where $Y_{\omega}$ is an anti-symmetric matrix, $X_S$ is a symmetric matrix, 
and $m_{\ell}\equiv \mathrm{diag}(m_e, m_{\mu}, m_{\tau})$ is a diagonal matrix with 
charged lepton masses. 
In general, this class of models contains a new singlet singly charged scalar $\omega^{\pm}$, which can have flavour anti-symmetric Yukawa interactions with lepton doublets as 
\begin{equation}
\mathcal{L}=\frac{1}{2}(Y_{\omega})^{ij}\epsilon_{ab}\bar{\ell}^{\text{c}a}_{Li}\ell_{Lj}^{b}\omega^++\text{h.c.}\;,
\end{equation}
where $\ell_{Li}$ denotes the left-handed SM lepton doublet field, 
$i,j=e,\mu,\tau$ are flavour indexes, the superscript $c$ denotes the charge conjugation and 
$a,b=1,2$ are SU(2) indexes. % chktex 36
Note that we take the basis that the charged leptons are mass eigenstates. 
The origin of the symmetric matrix $X_S$ depends on the detail of each model.

One can write the neutrino mass matrix $M_{\nu}$ 
in terms of the neutrino observables as 
\begin{equation}
M_{\nu}= U^*
\begin{pmatrix}
m_1&0&0\\
0&m_2&0\\
0&0&m_3	
\end{pmatrix}
U^{\dagger}\;,
\end{equation}
where $m_{1,2,3}$ are the mass eigenvalues of the light neutrinos and 
$U$ is the Pontecorvo-Maki-Nakagawa-Sakata (PMNS) matrix\cite{Pontecorvo:1957qd,Maki:1962mu}.
The PMNS matrix $U$ is parameterized as 
\begin{align}
&U=\nonumber \\
&\begin{pmatrix}
	c_{12}c_{13}&s_{12}c_{13}&s_{13}e^{-i\delta}\\
	-s_{12}c_{23}-c_{12}s_{13}s_{23}e^{i\delta}&
	c_{12}c_{23}-s_{12}s_{13}s_{23}e^{i\delta}&
	c_{13}s_{23}\\
	s_{12}s_{23}-c_{12}s_{13}c_{23}e^{i\delta}&
	-c_{12}s_{23}-s_{12}s_{13}c_{23}e^{i\delta}&
	c_{13}c_{23}
\end{pmatrix}
\begin{pmatrix}
e^{i\eta_1}&0&0 \\
0&e^{i\eta_2}&0 \\
0&0&1	
\end{pmatrix}\;,
\end{align}
with the abbreviations $c_{ij}=\cos\theta_{ij}$ and $s_{ij}=\sin\theta_{ij}$, and $\delta$ being the Dirac CP phase.
Eq.~(\ref{FlavorStructure}) provides $\det(M_{\nu})=0$, which means 
that the lightest neutrino is massless so that either $m_1$ or $m_3$ is zero.
Because of this, the Majorana phase $\eta_1$ can always be eliminated by the phase redefinition of the neutrino fields, and 
we set $\eta_1 = 0$ in the following. 
The cases with $m_1=0$ and $m_3=0$ correspond to the normal ordering (NO) and the inverted ordering (IO) neutrino masses.
Tab.~\ref{NeutrinoParam} shows the values of the neutrino parameters obtained from the global analysis of the current neutrino data. 

\begin{table}[ht]
\caption{The values of the neutrino parameters obtained from the global analysis. 
The numbers are taken from Ref.~\cite{Esteban:2020cvm} (the analysis without the Super-Kamiokande atmospheric data).\label{NeutrinoParam}}	
\begin{center}
\begin{tabular}{c||c|c||c|c}\hline
	&\multicolumn{2}{c||}{Normal Ordering}&
	\multicolumn{2}{c}{Inverted Ordering}\\ \hline
	&Best Fit& 3$\sigma$ range&Best Fit& 3$\sigma$ range\\ \hline
	$\theta_{12}/^\circ$&33.44&31.27 -- 35.86&33.45&31.27 -- 35.87\\ \hline
	$\theta_{23}/^\circ$&49.0&39.6 -- 51.8&49.3&39.9 -- 52.0\\ \hline
	$\theta_{13}/^\circ$&8.57&8.20 -- 8.97&8.61&8.24 -- 8.98\\ \hline
	$\delta/^\circ$&195&107 -- 403&286&192 -- 360\\ \hline
	$\Delta m_{21}^2/10^{-5}~\text{eV}^2$& 7.42&6.82 -- 8.04&7.42&6.82 -- 8.04\\ \hline 
	$\Delta m_{3\ell}^2/10^{-3}~\text{eV}^2$& 2.514&2.431--2.598&$-2.497$&$-2.583$ -- $-2.412$\\ \hline 
\end{tabular}
\end{center}
\end{table}

To explore the property of the flavour structure in Eq.~(\ref{FlavorStructure}), we parameterise $Y_{\omega}$ and $X_S$ as 
\begin{equation}
Y_{\omega}=
\begin{pmatrix}
0&f_{12}&f_{13}\\
-f_{12}&0&f_{23}\\
-f_{13}&-f_{23}&0	
\end{pmatrix}\;,\quad 
X_S=\begin{pmatrix}
X_{11}&X_{12}&X_{13}\\
X_{12}&X_{22}&X_{23}\\
X_{13}&X_{23}&X_{33}	
\end{pmatrix}\;.
\end{equation}
Since $m_e\ll m_{\mu}\ll m_{\tau}$ is satisfied, 
we can neglect the terms with $m_e$ in $M$ so that 
the elements $X_{1i}(i=1,2,3)$ become irrelevant to 
the neutrino mass matrix. 
This approximation leads to 
\begin{align}
M= Y_{\omega}m_{\ell}X_Sm_{\ell}Y_{\omega}^T
=
\begin{pmatrix}
*&*&*\\
*&m_{\tau}^2f_{23}^2X_{33}&-m_{\mu}m_{\tau}f_{23}^2X_{23}\\
*&-m_{\mu}m_{\tau}f_{23}^2X_{23}&m_{\mu}^2f_{23}^2X_{22}
\end{pmatrix}\;.
\label{FlavourStructure2}
\end{align}
for the right lower $2\times 2$ part of $M$. 
Within the 3$\sigma$ parameter range given in Tab.~\ref{NeutrinoParam}, 
$|M_{22}|\sim |M_{23}|
\sim|M_{33}|\sim \mathcal{O}(0.01)~\text{eV}$ 
is satisfied in most area of the allowed parameter space for both the NO and the IO masses\footnote{In certain points of the IO case, either $M_{22}$ or $M_{33}$ becomes tiny.}.
Therefore, it is reasonable to parameterise $X_{22}$ and $X_{33}$ as 
\begin{align}
X_{33} = \alpha \frac{m_{\mu}}{m_{\tau}}X_{23}
\;,\quad 
X_{22} = \beta \frac{m_{\tau}}{m_{\mu}}X_{23}
%	X_{33} =  \frac{m_{\mu}^2}{m_{\tau}^2} \alpha X_{22}\;,\quad 
%	X_{23} = \frac{m_{\mu}}{m_{\tau}}\beta X_{22}\;,\quad 
	\;,
\end{align}
with two complex parameters $\alpha$ and $\beta$ whose absolute values are 
of the order of one. 
%With this parametrisation, the $M_{11}$, $M_{12}$, and $M_{13}$ are given by 
%\begin{align}
%M_{11}
%\simeq  &\ 
%m_{\mu}m_{\tau}X_{23}(\beta f_{12}^2
%+2f_{12}f_{13}+\alpha f_{13}^2)\;,\nonumber\\
%M_{12}
%\simeq &\ 
%m_{\mu}m_{\tau}X_{23}(\alpha f_{13}f_{23}+\beta f_{12}f_{23})
%\;,\nonumber\\
%M_{13}
%\simeq  &\ 
%-m_{\mu}m_{\tau}X_{23}(\beta f_{12}f_{23}+ f_{13}f_{23})\;,
%M_{11}
%= &\ 
%m_{\mu}^2X_{22}(f_{12}^2+2\beta f_{12}f_{13}+\alpha f_{13}^2)\;,\nonumber\\
%M_{12}
%= &\ 
%m_{\mu}^2X_{22}(\alpha f_{13}f_{23}+\beta f_{12}f_{23})
%+\mathcal{O}(m_em_{\tau})\;,\nonumber\\
%M_{13}
%= &\ 
%-m_{\mu}^2X_{22}(f_{12}f_{23}+\beta f_{13}f_{23})
%+\mathcal{O}(m_em_{\tau})\;,\nonumber\\
%\end{align}
By introducing $k$ and $k'$ as 
\begin{align}
	f_{12}=kf_{23}\;,\quad 
	f_{13}=k'f_{23}\;,
\end{align}
we can write 
the neutrino mass matrix as
\begin{align}
M_{\nu}\propto M
=X_{23}f_{23}^2m_{\mu}m_{\tau}
\begin{pmatrix}
	\beta k^2+2 k k' + \alpha k^{\prime 2}&k + \alpha k'&-\beta k-k'\\
	k + \alpha k'&\alpha&-1\\
	-\beta k-k'&-1&\beta
\end{pmatrix}\;, 
\end{align}
which leads to 
\begin{align}
&\alpha=-\frac{(M_{\nu})_{22}}{(M_{\nu}){}_{23}}\;,\quad
\beta=-\frac{(M_{\nu})_{33}}{(M_{\nu}){}_{23}}\;,\quad
\label{eq:ZB_param1}
\\
&
k = \frac{1}{\alpha\beta-1}\left(\frac{(M_{\nu})_{12}}{(M_{\nu})_{23}}+\alpha\frac{(M_{\nu})_{13}}{(M_{\nu})_{23}}\right)\;,\quad
k' = -\frac{1}{\alpha\beta -1}\left(\beta\frac{(M_{\nu})_{12}}{(M_{\nu})_{23}}+\frac{(M_{\nu})_{13}}{(M_{\nu})_{23}}\right)\;.
%&\alpha=\frac{(M_{\nu})_{22}}{(M_{\nu})_{33}}\;,\quad
%\beta=-\frac{(M_{\nu})_{23}}{(M_{\nu})_{33}}\;,\quad
%\label{eq:ZB_param1}
%\\
%&
%k = -\frac{1}{\alpha-\beta^2}\left(\beta\frac{(M_{\nu})_{12}}{(M_{\nu})_{33}}+\alpha\frac{(M_{\nu})_{13}}{(M_{\nu})_{33}}\right)\;,\quad
%k' = \frac{1}{\alpha-\beta^2}\left(\frac{(M_{\nu})_{12}}{(M_{\nu})_{33}}+\beta\frac{(M_{\nu})_{13}}{(M_{\nu})_{33}}\right)\;.
\label{eq:ZB_param2}
\end{align}
Therefore, all the relevant parameters besides the overall factor $X_{22}f_{23}^2$ are determined by the neutrino parameters. 
Since the experimental 3$\sigma$ range of $\delta$ is still wide, and 
there is no restriction on $\eta_1$ and $\eta_2$, the parameters 
$\alpha$, $\beta$, $k$, and $k'$ vary in a certain range shown in Tab.~\ref{RangeOfParams}.

\begin{table}[ht]
\caption{The allowed range of the parameters $\alpha$, $\beta$, $k$, and $k'$, when 
the neutrino parameters are scanned in the 3$\sigma$ range given in Tab.~\ref{NeutrinoParam}.\label{RangeOfParams}}
\begin{center}
\begin{tabular}{c||c|c}\\ \hline
&Normal Ordering& Inverted Ordering\\ \hline
$|\alpha|$&$0.57\leq |\alpha|\leq 1.6$ & $0.0\leq |\alpha|\leq 2.4$\\ \hline
$\arg(\alpha)$& $-\pi\leq \arg(\alpha)\leq \pi$&$-\pi\leq \arg(\alpha)\leq \pi$ \\ \hline
$|\beta|$&$0.52\leq |\beta|\leq 1.6$  & $0.0\leq |\beta|\leq 2.8$\\ \hline
$\arg(\beta)$&$-\pi\leq \arg(\beta)\leq \pi$& $-\pi\leq \arg(\beta)\leq \pi$\\ \hline
$|k|$&$0.27\leq k \leq 0.67$ & $3.9\leq |k|\leq 5.3$\\ \hline
$\arg(k)$&$-0.27\leq \arg(k)\leq 0.32$ & $-2.9\leq \arg(k)\leq 0$\\ \hline
$|k'|$&$0.26\leq k'\leq 0.66$& $4.0\leq |k'|\leq 5.4$\\ \hline
$\arg(k')$&$-0.33\leq \arg(k')\leq 0.33$ & $0.21\leq \arg(k')\leq \pi$\\ \hline
\end{tabular}
\end{center}
\end{table}

\begin{figure}[ht]
\begin{center}
\includegraphics{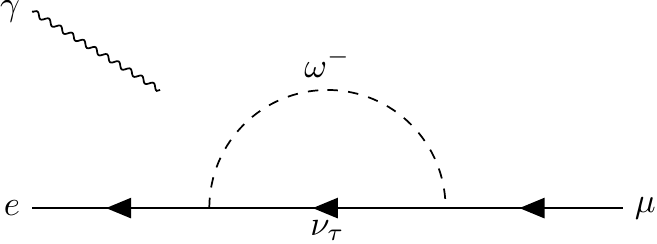}	
\end{center}
\caption{One loop diagram with $\omega$ contributing to $\mu\to e\gamma$ at the one-loop level. }\label{diag:meg}
\end{figure}

In this class of models, the singlet singly charged particle contributes to 
lepton flavour violation processes such as $\mu\to e\gamma$ via loop diagrams 
shown in Fig.~\ref{diag:meg}.
When this contribution is dominant, the branching ratio of $\mu\to e\gamma$ is proportional 
to $|f_{13}^*f_{23}|^2=|k^{\prime}|^2|f_{23}|^4$. It may imply an upper bound on $|k'|$.
A concrete example of the constraint from this issue is shown in the next section.

\section{Lepton flavour violation in the Zee-Babu model}
\begin{figure}[ht]
\begin{center}
\includegraphics{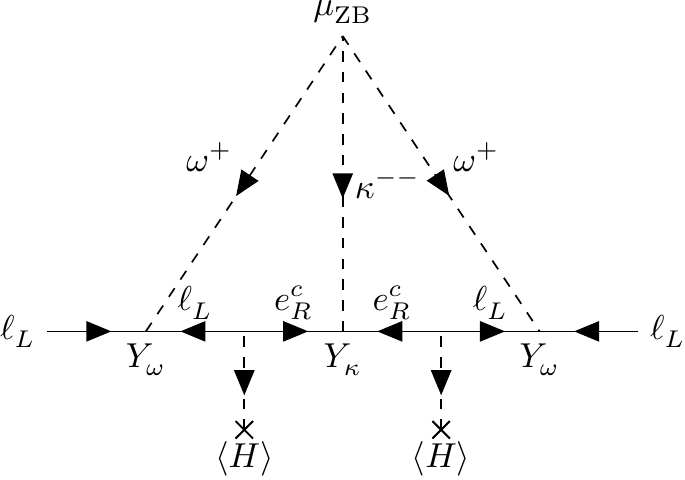}
\end{center}
\caption{The diagram relevant to the neutrino mass matrix in the Zee-Babu model.\label{Fig:MnuInZeeBabumodel}}
\end{figure}

The Zee-Babu model~\cite{Zee:1985id, Babu:1988ki} is a famous concrete example of the class of models discussed 
in the previous section. 
In this model, a singlet doubly charged scalar $\kappa^{--}$ is introduced in addition to $\omega^{-}$.
The relevant part of the Lagrangian is given by 
\begin{align}
\mathcal{L}_{\text{ZB}}=&
-\sum_{i,j=1}^3(Y_{\omega})^{ij}\bar{\ell}_{Li}^{c}\cdot \ell_{Lj}^{}\omega^+
-\sum_{i,j=1}^3(Y_{\kappa})^{ij}\bar{e}_{Ri}^{[e]}e_{Rj}^{c}\kappa^{--}
-\mu_{\text{ZB}}\omega^-\omega^-\kappa^{++}+\text{h.c.}\;.
\end{align}
The neutrino mass matrix is generated at the two-loop level through the diagram shown in Fig.~\ref{Fig:MnuInZeeBabumodel}.
Each element of the induced neutrino mass matrix 
is given by 
\begin{align}
(M_{\nu})^{ij}
=	
\left(\frac{1}{16\pi^2}\right)^2
\sum_{k,l=1}^3\frac{16\mu_{\text{ZB}}}{M_{\kappa}^2}
(Y_{\omega})^{ik}m_{e_k}(Y_{\kappa})^{kl}m_{e_l}(Y_{\omega})^{jl}
I(M_{\omega}^2,0|
M_{\omega}^2,0|M_{\kappa}^2)\;,
\label{MnuInZB}
\end{align}
where $M_{\omega}^2$ and 
$M_{\kappa}^2$ are the physical mass eigenvalues of the extra scalars,
$\omega^-$ and $\kappa^{--}$.
The loop function $I(m_{11}^2,m_{12}^2|m_{12}^2,m_{22}^2|M^2)$ is given by~\cite{vanderBij:1983bw,McDonald:2003zj}, 
\begin{align}
&I(m_{11}^2,m_{12}^2|m_{12}^2,m_{22}^2|M^2)\nonumber \\
&= 
\frac{1}{\pi^4}
\int d^4p\int d^4q
\frac{M^2}{(p^2+m_{11}^2)
(p^2+m_{12}^2)(q^2+m_{21}^2)(q^2+m_{22}^2)((p+q)^2+M^2)}\;.
%&=\ 
%\frac{M^4\left\{I(m_{12}^2|m_{22}^2|M^2)
%-I(m_{11}^2|m_{22}^2|M^2)
%-I(m_{12}^2|m_{21}^2|M^2)
%+I(m_{11}^2|m_{21}^2|M^2)\right\}}
%{(m_{11}^2-m_{12}^2)(m_{21}^2-m_{22}^2)}\;.
\end{align}
%The function $I(m_1^2|m_2^2|M^2)$ is defined as  
%\begin{equation}
%I(m_1^2|m_2^2|M^2)
%=-\left\{
%\frac{m_1^2}{M^2}f\left(\frac{m_2^2}{m_1^2},\frac{M^2}{m_1^2}\right)
%+\frac{m_2^2}{M^2}f\left(\frac{m_1^2}{m_2^2},\frac{M^2}{m_2^2}\right)
%+f\left(\frac{m_1^2}{M^2},\frac{m_2^2}{M^2}\right)
%\right\}\;,
%\end{equation}
%where 
%the function $f$ is given by 
%\begin{align}
%f(x,y)=&\ -\frac{1}{2}\log x \log y-\frac{1}{2}
%\left(\frac{x+y-1}{D}\right)\nonumber\\
%&\ \times 
%\left\{
%\mathrm{Li}_2\left(\frac{-\sigma_-}{\tau_+}\right)
%+\mathrm{Li}_2\left(\frac{-\tau_-}{\sigma_+}\right)
%-\mathrm{Li}_2\left(\frac{-\sigma_+}{\tau_-}\right)
%+\mathrm{Li}_2\left(\frac{-\tau_+}{\sigma_-}\right)
%\right.
%\nonumber\\
%&\left.\phantom{Spa}
%+\mathrm{Li}_2\left(\frac{y-x}{\sigma_-}\right)
%+\mathrm{Li}_2\left(\frac{x-y}{\tau_-}\right)
%-\mathrm{Li}_2\left(\frac{y-x}{\sigma_+}\right)
%-\mathrm{Li}_2\left(\frac{x-y}{\tau_+}\right)
%\right\}\;,
%\end{align}
%with 
%\begin{align}
%&D=\sqrt{1-2(x+y)+(x-y)^2}\;,\nonumber\\
%&\sigma_+=\frac{1}{2}\left(1-x+y+D\right)\;,\quad
%\tau_+=\frac{1}{2}\left(1+x-y+D\right)\;,\nonumber\\
%&\sigma_-=\frac{1}{2}\left(1-x+y-D\right)\;,\quad
%\tau_-=\frac{1}{2}\left(1+x-y-D\right)\;.
%\end{align}
In the approximation with $m_{e_i}=0$, the loop function can be evaluated as~\cite{Babu:2002uu,AristizabalSierra:2006gb}, 
\begin{align}
I(M_{\omega}^2,0|M_{\omega}^2,0|M_{\kappa}^2)
=&\ -\int_0^1dx\int_0^{1-x}dy
\frac{r}{x+(r-1)y+y^2}\log\frac{y(1-y)}{x+ry}
\nonumber \\
\simeq &\ 
\begin{cases}
2.8 r^2(r+0.31)^{-1.5}\;, &(r\gtrsim 1)\;,\\
1.98r(r+0.12)^{-0.23}\;, &(r< 1)\;,
\end{cases}
\end{align}
where $r=M_{\kappa}^2/M_{\omega}^2$.
The symmetric matrix $X_S$ in Eq.~(\ref{FlavorStructure}) can be identified with the 
Yukawa coupling matrix, $Y_{\kappa}$. 
Therefore, the relevant part of $Y_{\kappa}$ and $Y_{\omega}$ can be parameterized in the way shown in the previous section. 

In the Zee-Babu model, 
there are mainly two types of contributions to the LFV processes.
One is the singlet doubly charged scalar exchange at the tree level, 
which causes the LFV three body decays as $e_i^-\to e_j^-e_k^-e_l^+$. 
The decay width is given by 
\begin{align}
\Gamma(e_i^-\to e_j^-e_k^-e_l^+)
=\frac{C_{jk}}{8}\frac{m_{e_i}^5}{192\pi^3}
\left|\frac{(Y_{\kappa})^{il}_{}(Y_{\kappa})^{*jk}_{}}{M_{\kappa}^2}\right|^2\;,
\end{align}
where 
$C_{jk}$ denotes a statistical factor as
\begin{equation}
C_{jk}=\begin{cases}
1& (j=k)\\
2& (j\neq k)	
\end{cases}\;.
\end{equation}
Another is a one-loop contribution to $e_i\to e_j\gamma$ via the same type of 
the diagram shown in Fig.~\ref{diag:meg}.  
The decay width of $e_i\to e_j\gamma$ is given by 
\begin{align}
\Gamma(e_i\to e_j+\gamma)=\frac{\alpha_e}{4}m_{e_i}^5\left(|A_L^{ji}|^2+|A_R^{ji}|^2\right)\;,	
\end{align}
with $\alpha_e = e^2/(4\pi)$ and 
\begin{align}
A_L^{ji}\simeq &\ -\frac{1}{(4\pi){}^2}
\frac{(Y_{\omega})^{*kj}_{}(Y_{\omega}){}^{ki}}{3M_{\omega}^2}\;,\quad
A_R^{ji}= 
-\frac{1}{(4\pi)^2}
\frac{4(Y_{\kappa})^{*kj}_{}(Y_{\kappa}){}^{ki}}{3M_{\kappa}^2}\;.
\end{align}
As discussed in the previous section, the elements $(Y_{\kappa}){}_{i1}$ are
irrelevant to the neutrino mass matrix. In the following, we fix 
$(Y_{\kappa})_{i1}=0$ for simplicity.
If non-zero $(Y_{\kappa})_{i1}$ is taken into account, 
$\text{Br}(\mu\to e\gamma)$ becomes enhanced. 

\begin{figure}[ht]
\begin{center}
\includegraphics[width=7cm]{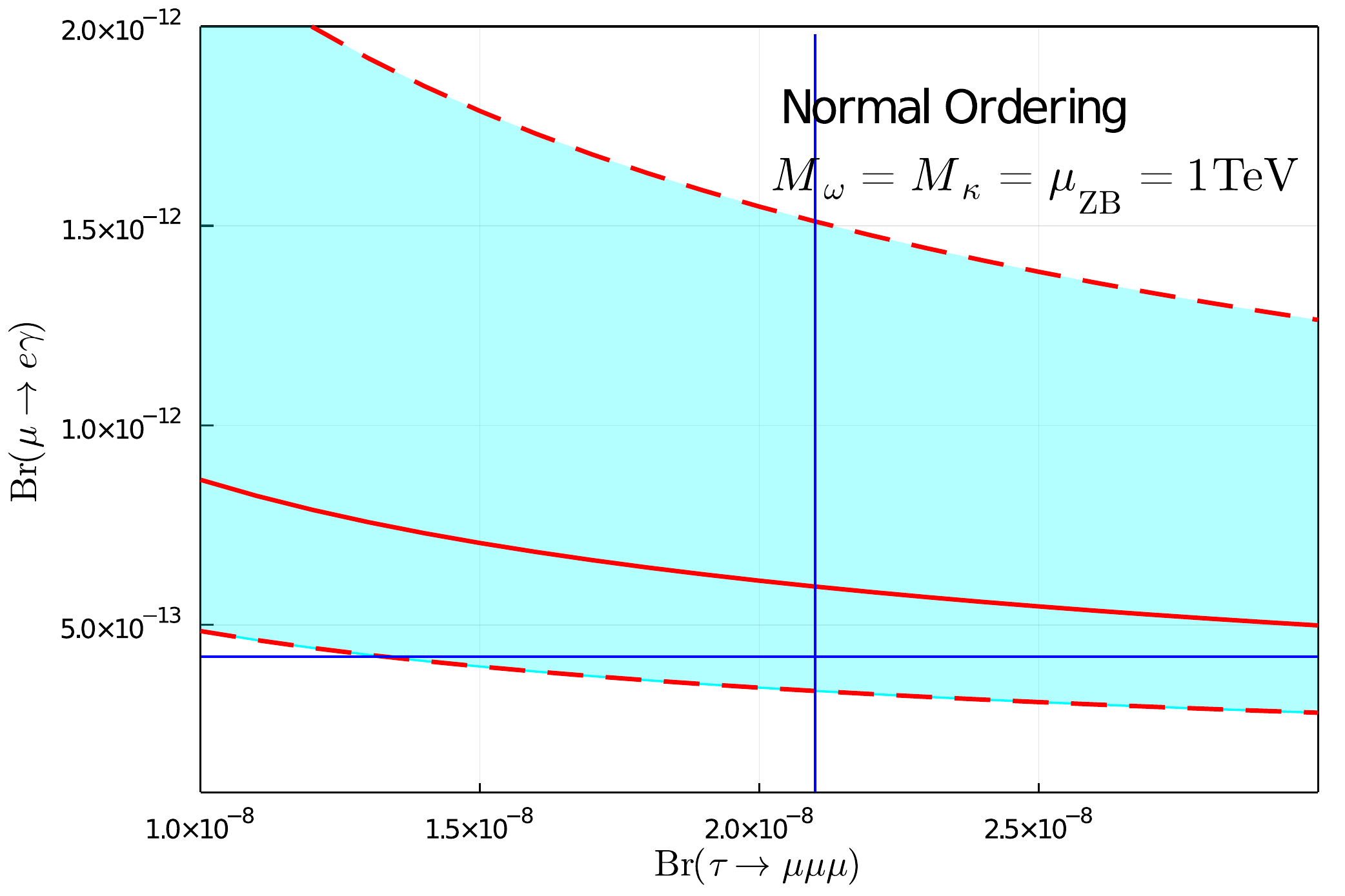}
\end{center}
\caption{The correlation between $\text{Br}(\tau\to \mu\mu\mu)$ and 
the minimal value of $\text{Br}(\mu\to e\gamma)$ in the NO case is shown. 
The massive model parameters are fixed as $M_{\omega}=M_{\kappa}=\mu_{\text{ZB}}=1.0~\text{TeV}$.
The shaded region is obtained by scanning the other oscillation parameters in the 3$\sigma$ range 
shown in Tab.~\ref{NeutrinoParam} and scanning $\eta_2$ in $0\leq \eta_2\leq 2\pi$. 
On the solid red curve, the neutrino oscillation parameters 
are fixed to be the best fit values listed in Tab.~\ref{NeutrinoParam} and 
$\eta_2=0$. 
The horizontal and vertical blue solid lines show the current upper bound on 
$\text{Br}(\mu\to e\gamma)$ and 
$\text{Br}(\tau\to \mu\mu\mu)$.
}
\label{BrMEGvsBrT2MMM_NO}
\end{figure}

Let us first consider the NO neutrino case. 
In this case, $(Y_{\kappa}){}_{22}\neq 0$ is satisfied within the 3$\sigma$ range of neutrino parameters listed in Tab.~\ref{NeutrinoParam}. 
Once the neutrino parameters $m_i$, $\theta_{ij}$, $\delta$, $\eta_1$, and $\eta_2$, and 
massive parameters $M_{\omega}$, $M_{\kappa}$, and $\mu_{\text{ZB}}$ are fixed, 
the element $(M_{\nu})_{23}$ determines the size of $f_{23}^2Y_{23}$.
It leads to a strong correlation $\text{Br}(\mu\to e\gamma){}^2\propto \text{Br}(\tau\to \mu\mu\mu)$.
This correlation is displayed in Fig.~\ref{BrMEGvsBrT2MMM_NO}.
One can see that $\text{Br}(\mu\to e\gamma)$ is minimized when $\text{Br}(\tau\to \mu\mu\mu)$ is maximized as 
$\text{Br}(\tau\to \mu\mu\mu)=2.1\times 10^{-8}$ which is the upper limit from the experiment~\cite{Hayasaka:2010np}. 

\begin{figure}[ht]
\begin{center}
\includegraphics[width=7cm]{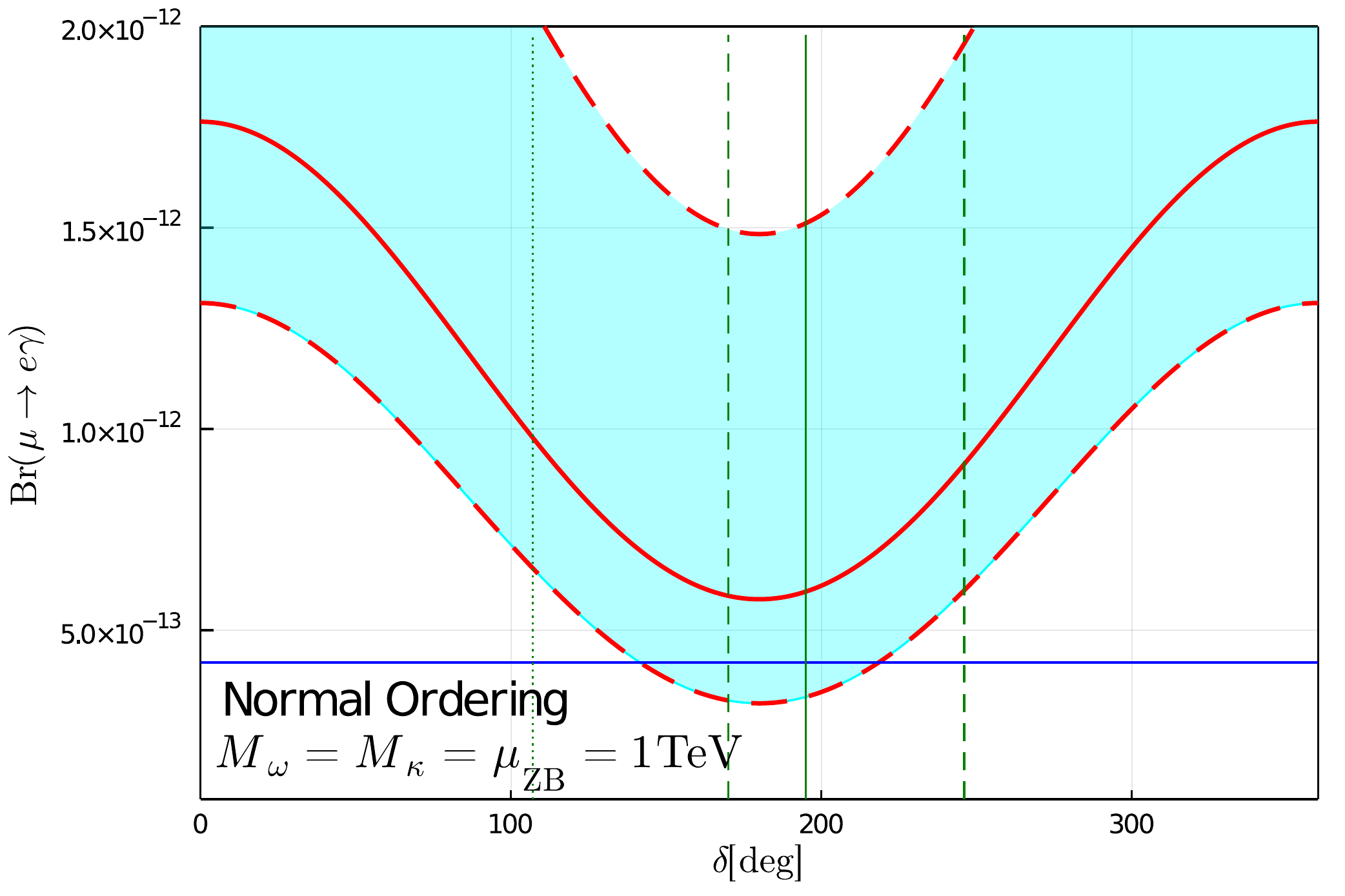}
\end{center}
\caption{
The dependence of the minimal value of $\text{Br}(\mu\to e\gamma)$ on 
the Dirac CP phase in the PMNS matrix, $\delta$.
$M_{\omega}=M_{\kappa}=\mu_{\text{ZB}}=1.0~\text{TeV}$.
The shaded region is obtained by scanning the other oscillation parameters 
except for $\delta$ in the 3$\sigma$ range 
shown in Tab.~\ref{NeutrinoParam} and scanning $\eta_2$ in $0\leq \eta_2\leq 2\pi$. 
On the red solid curve, the neutrino oscillation parameters 
are fixed to be the best fit values listed in Tab.~\ref{NeutrinoParam} and 
$\eta_2=0$. 
The $(Y_{\kappa})_{23}$ is determined to satisfy 
$\text{Br}(\tau\to \mu\mu\mu)=2.1\times 10^{-8}$. 
The horizontal solid line shows the upper bound on $\text{Br}(\mu\to e\gamma)$.
The vertical solid line denotes the best fit value of $\delta$. 
The dashed and dotted line are the boundary of the $1\sigma$ and 
$3\sigma$ allowed range of $\delta$, respectively.}
\label{delCP-BrMEG-NO}
\end{figure}

In Fig.~\ref{delCP-BrMEG-NO}, we show the dependence of the minimal value of $\text{Br}(\mu\to e\gamma)$ on 
the Dirac CP phase in the PMNS matrix, $\delta$. 
The other oscillation parameters are scanned in the 3$\sigma$ range given in Tab.~\ref{NeutrinoParam},
the Majorana CP phase $\eta_2$ is scanned in the range $0\leq \eta_2\leq 2\pi$, and 
we fix the massive parameters in the extra scalar sector as 
$M_{\omega}=M_{\kappa}=\mu_{\text{ZB}}=1.0~\text{TeV}$.
We take the maximal value of $|Y_{23}|$ which leads to $\text{Br}(\tau \to \mu\mu\mu)=2.1\times 10^{-8}$, 
in order to minimize $|f_{23}|$ and $\text{Br}(\mu\to e\gamma)$. 

\begin{figure}[ht]
\begin{center}
\includegraphics[width=10cm]{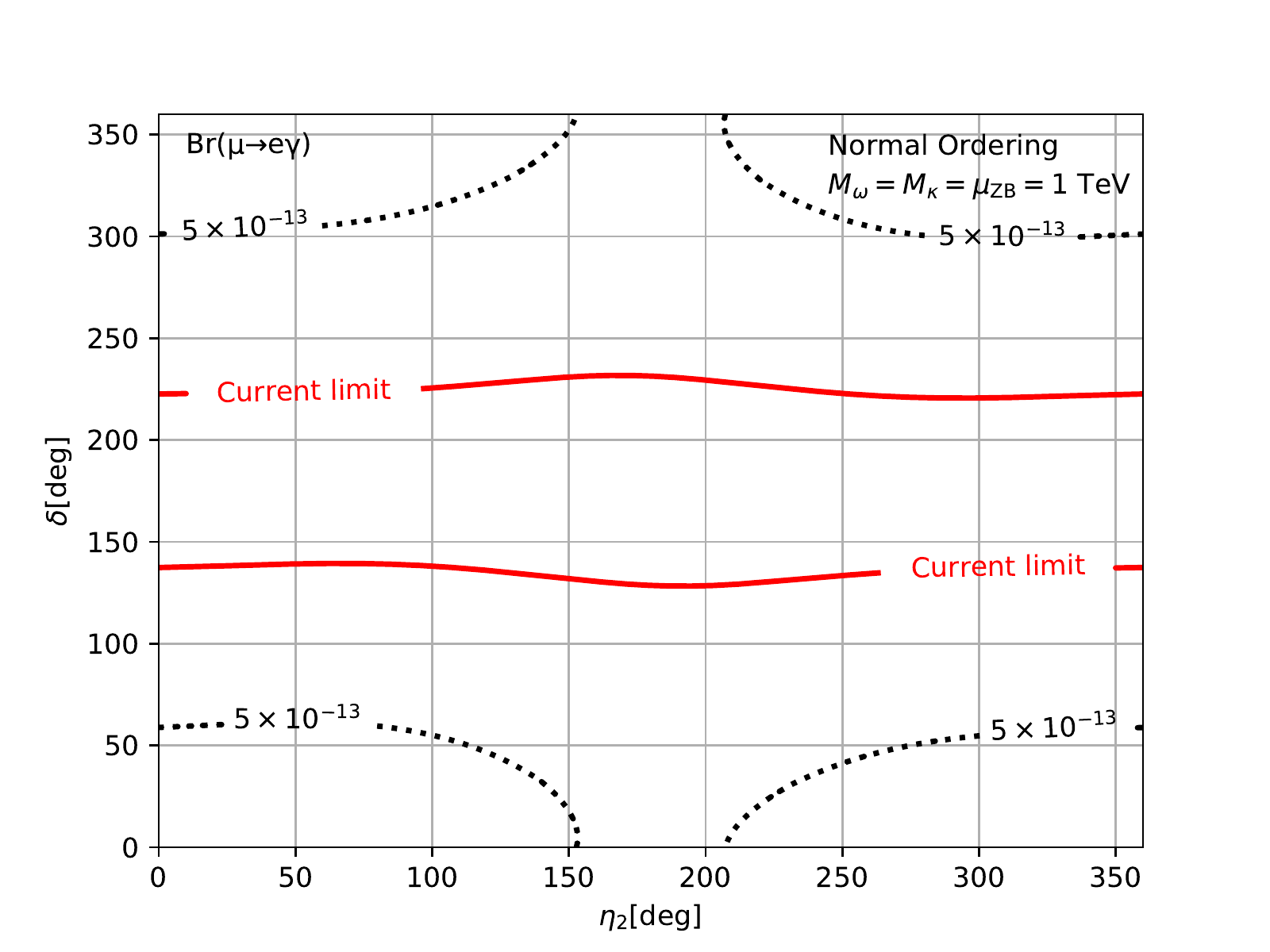}
\end{center}
\caption{
The contour of the minimal value of $\text{Br}(\mu\to e\gamma)$ on the $\eta_2$-$\delta$ plane.
The other neutrino oscillation parameters are scanned 
in the 3$\sigma$ range 
shown in Tab.~\ref{NeutrinoParam}.
We fix the massive parameters in the extra scalar sector as 
$M_{\omega}=M_{\kappa}=\mu_{\text{ZB}}=1.0~\text{TeV}$.
The solid curves corresponds to the current upper bound $\text{Br}(\mu\to e\gamma)=4.2\times 10^{-13}$.
}
\label{eta2vsdelCP_NO}
\end{figure}

Fig.~\ref{eta2vsdelCP_NO} shows the contour of $\text{Br}(\mu\to e\gamma)$ on the parameter plane of $\eta_2$ and $\delta$. The other neutrino parameters are scanned within the 3$\sigma$ range.
The solid red curves correspond to the current upper bound $\text{Br}(\mu\to e\gamma)=4.2\times 10^{-13}$~\cite{TheMEG:2016wtm}, and the area between these two curves is the allowed region. 


\begin{figure}[ht]
\begin{center}
	\begin{tabular}{cc}
	\includegraphics[width=6cm]{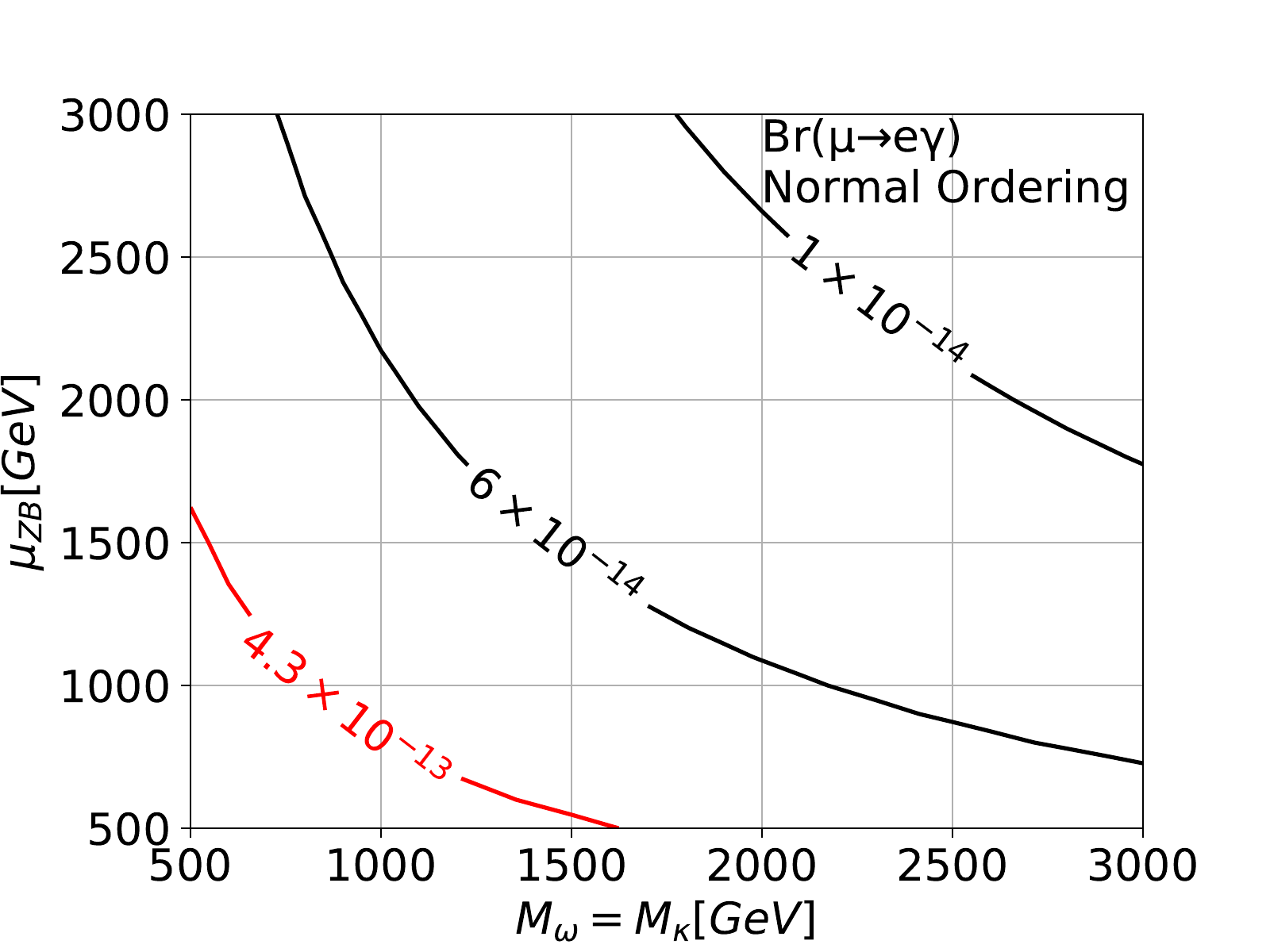}&
	\includegraphics[width=6cm]{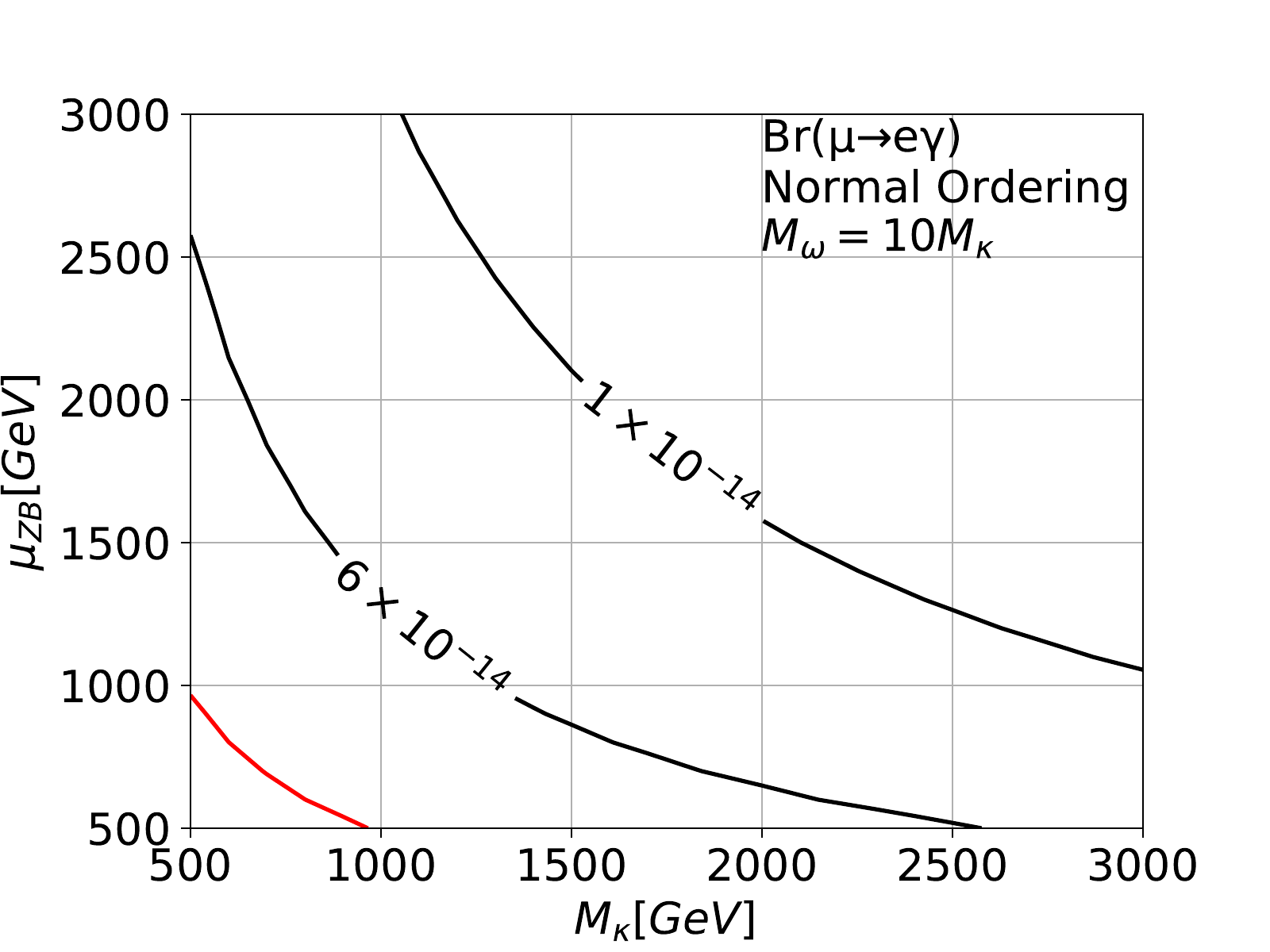}\\
	(a) $M_{\omega}=M_{\kappa}$& (b) $M_{\omega}=10M_{\kappa}$\\
	\includegraphics[width=6cm]{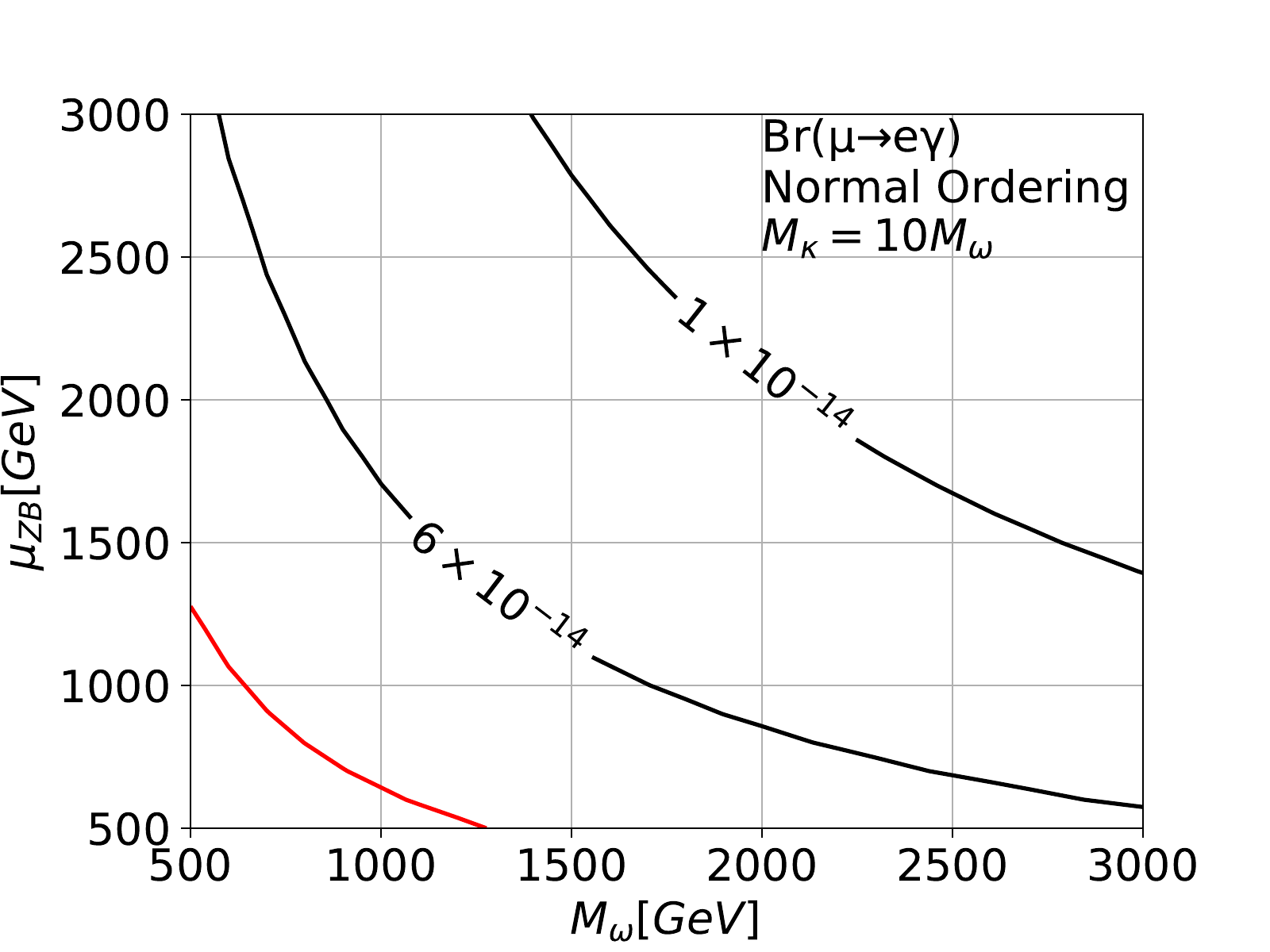} &\\
	(c) $M_{\kappa}=10M_{\omega}$&
	\end{tabular}
\end{center}
\caption{
The contour of the minimal value of $\text{Br}(\mu\to e\gamma)$ in each point on the 
parameter plane of an extra scalar mass $M_{\omega}$ or $M_{\kappa}$ and $\mu_{\text{ZB}}$. 
Three patterns of the extra scalar masses are plotted as (a) $M_{\omega}=M_{\kappa}$, (b) $M_{\omega}=10M_{\kappa}$, 
and (c) $M_{\kappa}=10M_{\omega}$.
The lower-left side of the red curve in each plot is excluded by the bound of $\text{Br}(\mu\to e\gamma)$. 
The neutrino parameters are scattered within the $3\sigma$ range given in Tab.~\ref{NeutrinoParam}, and the Majorana CP phase $\eta_2$ is considered 
in $-\pi\leq \eta_2\leq \pi$.
}
\label{MandmuZBrange_NO}
\end{figure}

In Fig.~\ref{MandmuZBrange_NO}, we show the contour of the minimal value of 
$\text{Br}(\mu\to e\gamma)$ in each point on the parameter plane of a extra scalar mass $M_{\omega}$ 
or $M_{\kappa}$ and $\mu_{\text{ZB}}$.
The LHC bound on the extra scalar masses is studied in the literature~\cite{Schmidt:2014zoa,Herrero-Garcia:2014hfa,Alcaide:2017dcx}, and the region of $m_{\kappa}, m_{\omega}>500$~GeV is not constrained by the direct search of extra particles. 
The neutrino parameters are scanned within the $3\sigma$ range given in Tab.~\ref{NeutrinoParam}, and $\eta_2$ is considered in $-\pi\leq \eta_2\leq \pi$.
In the figures (a), (b), and (c), we consider the case with 
$M_{\omega}=M_{\kappa}$, $M_{\omega}=10M_{\kappa}$, and $M_{\kappa}=10M_{\omega}$, respectively. 
The left-below area of the solid red curve in each figure is excluded by the current experimental bound of $\text{Br}(\mu\to e\gamma)$. 
Since the $\text{Br}(\mu\to e\gamma)$ is dominated by the $\omega$ exchange diagram, 
the constraint is significantly relaxed for $M_{\omega}\gg M_{\kappa}$ case. 
On the other hand, the contribution to $\text{Br}(\tau\to \mu\mu\mu)$ is suppressed in $M_{\kappa}\gg M_{\omega}$, and the size of $|f_{23}|$ can be small. 
Therefore, the constraint is relaxed in comparison to the $M_{\omega}=M_{\kappa}$ case. 
Let us comment on the case with large $\mu_{\text{ZB}}$. If $\mu_{\text{ZB}}$ is too large compared to $M_{\omega}$ and $M_{\kappa}$, the electroweak vacuum might be unstable, so that such a case might be constrained. 
The quantitative study of vacuum stability is out of the scope of this paper. 
The sensitivity to $\text{Br}(\mu\to e\gamma)$ will go down 
to $6\times 10^{-14}$ at the MEG-II experiment~\cite{Baldini:2018nnn}, and a wider region of the parameter space will be explored.

\begin{figure}[ht]
	\begin{center}
		\begin{tabular}{cc}
			\includegraphics[width=6cm]{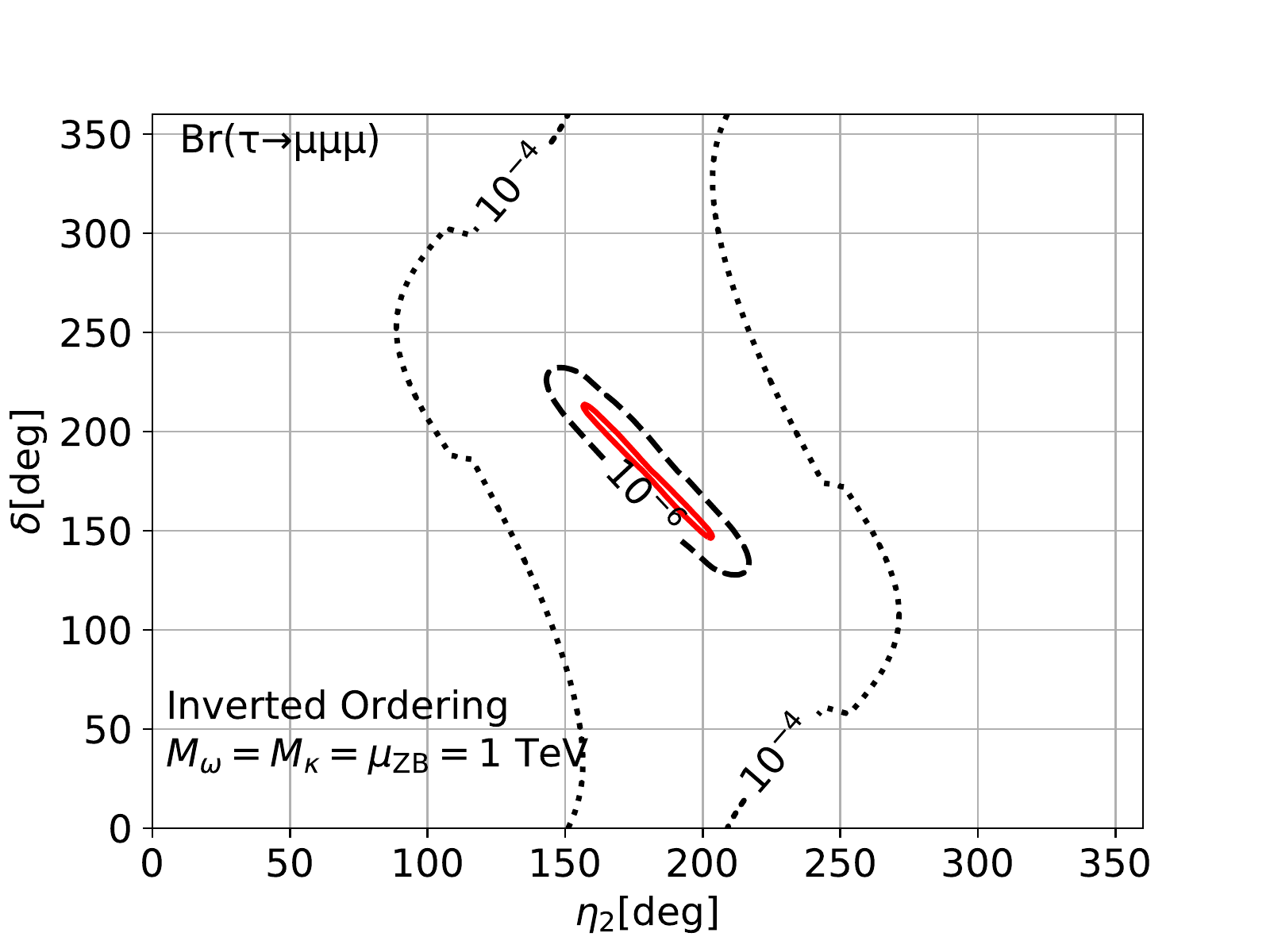}&
			\includegraphics[width=6cm]{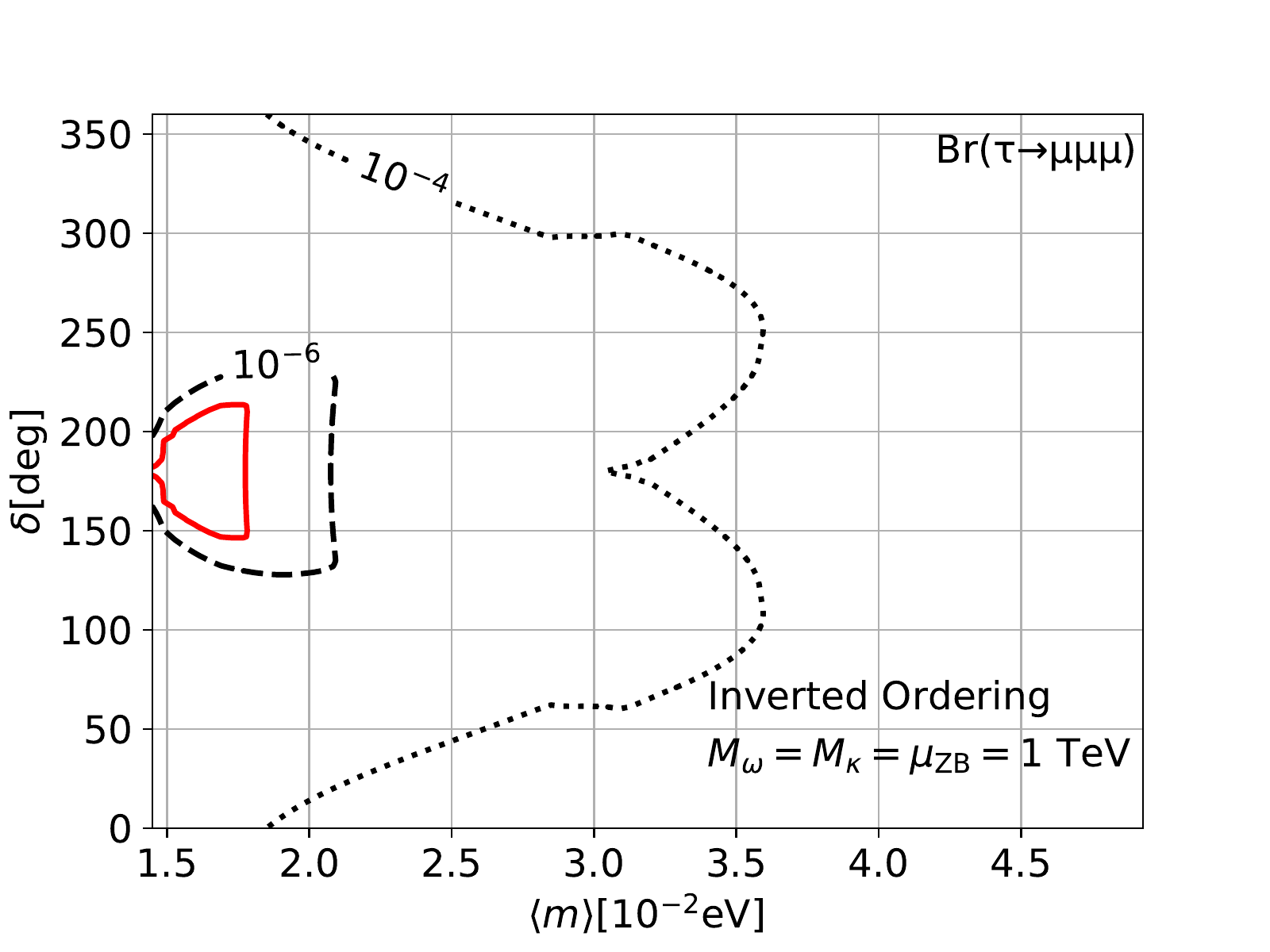}\\
			(a)&(b)
		\end{tabular}
	\end{center}
	\caption{
	The contour of the minimal value of $\text{Tr}(\tau\to\mu\mu\mu )$ 
	in the IO case are shown on (a) the $\eta_2$-$\delta$ plane and (b) the $\langle m\rangle$-$\delta$ plane. 
	The other neutrino oscillation parameters are scanned 
	in the 3$\sigma$ range 
	shown in Tab.~\ref{NeutrinoParam}.
	We fix the massive parameters in the extra scalar sector as 
	$M_{\omega}=M_{\kappa}=\mu_{\text{ZB}}=1.0~\text{TeV}$.
	The solid curves corresponds to the current upper bound $\text{Br}(\tau\to \mu\mu\mu)=2.1\times 10^{-8}$.
	}
	\label{eta2vsdelCP_IO}
	\end{figure}

In the IO case, $(Y_{\kappa})_{22}=0$ is realized in a certain set of the neutrino mixing parameters. 
In such parameter points, $\text{Br}(\tau\to \mu\mu\mu)$ satisfies 
the experimental limit with the rather large value of $(Y_{\kappa})_{23}$ witch makes $f_{23}$ small enough to satisfy 
$\text{Br}(\mu\to e\gamma)<4.3\times 10^{-13}$.
In Fig.~\ref{eta2vsdelCP_IO}, we show the contour plots
of $\text{Br}(\tau\to \mu\mu\mu)$ on the plane of 
$\eta_2$ and $\delta$, and on the plane of $\langle m\rangle$ and $\delta$.
Here, $\langle m\rangle$ is the effective Majorana mass parameter of the neutrinoless double beta decay, which is defined by 
\begin{equation}
	\langle m\rangle = \left|\sum_{i=1,2,3}m_iU_{ei}^2\right|\;.
\end{equation}
We scanned the other neutrino parameters within the 3$\sigma$ range given in Tab.~\ref{NeutrinoParam}, and we tune the $|f_{23}|$ to be $\text{Br}(\mu\to e\gamma)=4.3\times 10^{-13}$ which is the current upper limit. 
The red curves show the experimental bound of $\text{Br}(\tau\to \mu\mu\mu)$, and inside the curve is allowed.
By the strong constraint on the lepton flavour violation, the lower value of $\langle m\rangle$ is preferred.

\section{Summary}

We have discussed the lepton flavour violating processes and their phenomenological consequences in a class of models with radiative neutrino mass generation. 
We focus on the models where the induced neutrino mass matrix has a specific structure given by $M_{\nu}\propto Y_{\omega}m_{\ell}X_Sm_{\ell}Y_{\omega}^T$.
We have utilized an approximation that the electron mass in $m_{\ell}$ is neglected, 
and we have found the relations between the neutrino mass matrix and the Yukawa coupling matrices
of the extra scalars. 

We have applied such relations to the Zee-Babu model as a concrete example. 
We have found that the lepton flavour violation processes, $\tau\to \mu\mu\mu$ and $\mu\to e\gamma$, are powerful tools to investigate the Zee-Babu model. 
They constrain the neutrino parameters as well as the massive parameters in the extra scalar sector.

The MEG-II will search $\mu\to e\gamma$ down to $\text{Br}(\mu\to e\gamma)\leq 6\times 10^{-14}$~\cite{Baldini:2018nnn}, and
the expected sensitivity of the upper limit on $\text{Br}(\tau\to \mu\mu\mu)$ is $3.3\times 10^{-10}$ at the Belle-II experiment~\cite{Kou:2018nap}.
The neutrino parameters will be more precisely determined by the neutrino oscillation experiments. 
For example, if $\delta=0$ or $\delta=180^{\circ}$ is the case, the Dirac CP phase will be measured with the uncertainty of $7.2^{\circ}$ at the Hyper-Kamiokande experiment~\cite{Abe:2018uyc}.
The sensitivities of the other oscillation parameters will also be significantly improved. 

In the near future, the Zee-Babu model can be probed by multiple searches for the lepton flavour violation, neutrino experiments, and new particle search at high energy collider experiments.

\section*{Acknowledgments}
This work is supported in part by the Japan Society for the Promotion
of Science (JSPS) KAKENHI Grants No.~19K03860, No.~19K03865, No.~21H00060 (O.S.) and  20H00160 (T.S.).

\bibliography{ISS}

\begin{thebibliography}{10}

\bibitem{Minkowski:1977sc}
P. Minkowski,
\newblock Phys. Lett. 67B (1977) 421.

\bibitem{Yanagida:1979as}
T. Yanagida,
\newblock Conf. Proc. C 7902131 (1979) 95.

\bibitem{Yanagida:1980xy}
T. Yanagida,
\newblock Prog.Theor.Phys. 64 (1980) 1103.

\bibitem{GellMann:1980vs}
M. Gell-Mann, P. Ramond and R. Slansky,
\newblock Conf.Proc. C790927 (1979) 315, 1306.4669.

\bibitem{Mohapatra:1979ia}
R.N. Mohapatra and G. Senjanovic,
\newblock Phys. Rev. Lett. 44 (1980) 912.

\bibitem{Weinberg:1979sa}
S. Weinberg,
\newblock Phys. Rev. Lett. 43 (1979) 1566.

\bibitem{Konetschny:1977bn}
W. Konetschny and W. Kummer,
\newblock Phys. Lett. B 70 (1977) 433.

\bibitem{Cheng:1980qt}
T.P. Cheng and L.F. Li,
\newblock Phys. Rev. D 22 (1980) 2860.

\bibitem{Lazarides:1980nt}
G. Lazarides, Q. Shafi and C. Wetterich,
\newblock Nucl. Phys. B 181 (1981) 287.

\bibitem{Schechter:1980gr}
J. Schechter and J.W.F. Valle,
\newblock Phys. Rev. D 22 (1980) 2227.

\bibitem{Magg:1980ut}
M. Magg and C. Wetterich,
\newblock Phys. Lett. B 94 (1980) 61.

\bibitem{Mohapatra:1980yp}
R.N. Mohapatra and G. Senjanovic,
\newblock Phys. Rev. D 23 (1981) 165.

\bibitem{Zee:1980ai}
A. Zee,
\newblock Phys. Lett. 93B (1980) 389,
\newblock [Erratum: Phys. Lett.95B,461(1980)].

\bibitem{Cai:2017jrq}
Y. Cai et~al.,
\newblock Front. in Phys. 5 (2017) 63, 1706.08524.

\bibitem{Esteban:2020cvm}
I. Esteban et~al.,
\newblock JHEP 09 (2020) 178, 2007.14792.

\bibitem{Kanemura:2015cca}
S. Kanemura and H. Sugiyama,
\newblock Phys. Lett. B 753 (2016) 161, 1510.08726.

\bibitem{Kanemura:2016ixx}
S. Kanemura, K. Sakurai and H. Sugiyama,
\newblock Phys. Lett. B 758 (2016) 465, 1603.08679.

\bibitem{Zee:1985id}
A. Zee,
\newblock Nucl. Phys. B 264 (1986) 99.

\bibitem{Babu:1988ki}
K.S. Babu,
\newblock Phys. Lett. B 203 (1988) 132.

\bibitem{Krauss:2002px}
L.M. Krauss, S. Nasri and M. Trodden,
\newblock Phys. Rev. D 67 (2003) 085002, hep-ph/0210389.

\bibitem{Babu:2002uu}
K.S. Babu and C. Macesanu,
\newblock Phys. Rev. D 67 (2003) 073010, hep-ph/0212058.

\bibitem{AristizabalSierra:2006gb}
D. Aristizabal~Sierra and M. Hirsch,
\newblock JHEP 12 (2006) 052, hep-ph/0609307.

\bibitem{Nebot:2007bc}
M. Nebot et~al.,
\newblock Phys. Rev. D 77 (2008) 093013, 0711.0483.

\bibitem{Schmidt:2014zoa}
D. Schmidt, T. Schwetz and H. Zhang,
\newblock Nucl. Phys. B 885 (2014) 524, 1402.2251.

\bibitem{Herrero-Garcia:2014hfa}
J. Herrero-Garcia et~al.,
\newblock Nucl. Phys. B 885 (2014) 542, 1402.4491.

\bibitem{Alcaide:2017dcx}
J. Alcaide, M. Chala and A. Santamaria,
\newblock Phys. Lett. B 779 (2018) 107, 1710.05885.

\bibitem{Pontecorvo:1957qd}
B. Pontecorvo,
\newblock Zh. Eksp. Teor. Fiz. 34 (1957) 247.

\bibitem{Maki:1962mu}
Z. Maki, M. Nakagawa and S. Sakata,
\newblock Prog. Theor. Phys. 28 (1962) 870.

\bibitem{vanderBij:1983bw}
J. van~der Bij and M.J.G. Veltman,
\newblock Nucl. Phys. B 231 (1984) 205.

\bibitem{McDonald:2003zj}
K.L. McDonald and B.H.J. McKellar,
\newblock (2003), hep-ph/0309270.

\bibitem{Hayasaka:2010np}
K. Hayasaka et~al.,
\newblock Phys. Lett. B 687 (2010) 139, 1001.3221.

\bibitem{TheMEG:2016wtm}
MEG, A.M. Baldini et~al.,
\newblock Eur. Phys. J. C 76 (2016) 434, 1605.05081.

\bibitem{Baldini:2018nnn}
MEG II, A.M. Baldini et~al.,
\newblock Eur. Phys. J. C 78 (2018) 380, 1801.04688.

\bibitem{Kou:2018nap}
Belle-II, W. Altmannshofer et~al.,
\newblock PTEP 2019 (2019) 123C01, 1808.10567,
\newblock [Erratum: PTEP 2020, 029201 (2020)].

\bibitem{Abe:2018uyc}
Hyper-Kamiokande, K. Abe et~al.,
\newblock (2018), 1805.04163.

\end{thebibliography}
\end{document}